\documentclass[twocolumn,showpacs,preprintnumbers,amsmath,amssymb]{revtex4}
\usepackage{graphicx}

\begin{document}
\title{Quantum secret sharing based on Smolin states alone}
\author{Guang Ping He}
 \email{hegp@mail.sysu.edu.cn}
\affiliation{School of Physics \& Engineering and Advanced
Research Center, Sun Yat-sen University, Guangzhou 510275, China\\
and Center of Theoretical and Computational Physics, The University
of Hong Kong, Pokfulam Road, Hong Kong, China}
\author{Z. D. Wang}
 \email{zwang@hkucc.hku.hk}
\author{Yan-Kui Bai}
 \email{ykbai@semi.ac.cn}
\affiliation{Department of Physics and Center of Theoretical and
Computational Physics, The University of Hong Kong, Pokfulam Road,
Hong Kong, China}

\begin{abstract}
It was indicated [Yu 2007 Phys. Rev. A 75 066301] that a previous
proposed quantum secret sharing (QSS) protocol based on Smolin
states [Augusiak 2006 Phys. Rev. A 73 012318] is insecure against an
internal cheater. Here we build a different QSS protocol with Smolin
states alone, and prove it to be secure against known cheating
strategies. Thus we open a promising venue for building secure QSS
using merely Smolin states, which is a typical kind of bound
entangled states. We also propose a feasible scheme to implement the
protocol experimentally.
\end{abstract}

\pacs{03.67.Hk,03.67.Dd}
\maketitle

\newpage

\section{Introduction}

The properties of Smolin states \cite{qi474-23} have caught great interests
recently. It was shown \cite{qi474,qi356} that they can maximally violate
simple correlation Bell inequalities, and thus reduce communication
complexity. On the other hand, as a typical kind of bound entangled (i.e.,
cannot be distilled to pure entangled form with local operations and
classical communications (LOCC) ) states, Smolin states do not allow for
secure key distillation. This indicates that neither entanglement nor
maximal violation of Bell inequalities implies directly the presence of a
quantum secure key. Thus it becomes an intriguing question how useful Smolin
states can be for quantum cryptography. Especially, it was left an open
question in Refs. \cite{qi474,qi356} whether Smolin states can lead to
secure quantum secret sharing (QSS) \cite{qi103,qi46}. This question was
further indicated to be non-trivial by Ref. \cite{qi483}, in which an
explicit cheating strategy was proposed, showing that a class of QSS
protocols using Smolin states can be broken if one of the participants is
dishonest.

In this paper, a four-party QSS protocol based on Smolin states is proposed,
and proven to be secure against the cheating strategy proposed in Ref. \cite
{qi483} as well as other known attacks. A feasible scheme for implementing
our protocol experimentally is proposed. Building multi-party secure QSS
protocols on generalized Smolin states \cite{qi474} is also addressed. These
findings may help to answer the question whether Smolin states and other
bound entangled states can lead to secure QSS.

\section{The original protocol and the cheating strategy}

The original Smolin state is a mixed state of four qubits $A$, $B$, $C$ and $%
D$\ described by the density matrix
\begin{eqnarray}
\rho _{ABCD}^{S} &=&\frac{1}{4}(\left| \Phi ^{+}\right\rangle
_{AB}\left\langle \Phi ^{+}\right| \otimes \left| \Phi ^{+}\right\rangle
_{CD}\left\langle \Phi ^{+}\right|  \nonumber \\
&&+\left| \Phi ^{-}\right\rangle _{AB}\left\langle \Phi ^{-}\right| \otimes
\left| \Phi ^{-}\right\rangle _{CD}\left\langle \Phi ^{-}\right|  \nonumber
\\
&&+\left| \Psi ^{+}\right\rangle _{AB}\left\langle \Psi ^{+}\right| \otimes
\left| \Psi ^{+}\right\rangle _{CD}\left\langle \Psi ^{+}\right|  \nonumber
\\
&&+\left| \Psi ^{-}\right\rangle _{AB}\left\langle \Psi ^{-}\right| \otimes
\left| \Psi ^{-}\right\rangle _{CD}\left\langle \Psi ^{-}\right| ).
\label{Smolin}
\end{eqnarray}
Here $\left| \Phi ^{\pm }\right\rangle =(\left| 00\right\rangle \pm \left|
11\right\rangle )/\sqrt{2}$\ and $\left| \Psi ^{\pm }\right\rangle =(\left|
01\right\rangle \pm \left| 10\right\rangle )/\sqrt{2}$\ denote the four Bell
states.

Now consider the task of QSS among four parties Alice, Bob, Charlie
and Diana. The model of QSS studied in this paper includes the
following essential features. (I) The goal of the process is that
Alice, who has a classical secret bit to be shared, encodes the bit
with certain quantum states and sends them
to the other three parties, so that they can retrieve the secret bit \textit{%
if and only if} all the three of them collaborate. (II) In QSS, it
is generally assumed that Alice always acts honestly. That is, we
don't consider the case where Alice wants to cause the participants
to accept inconsistent versions of her secret bit. (III) A QSS
protocol is called secure if it can stand the following two types of
attack, (1) ``passive'' attacks, i. e., eavesdropping from external
attackers, and (2) ``active attacks'', i. e., one or some of the
legal participants trying to gain non-trivial amount of information
on the secret bit without the collaboration of all participants
(except Alice).

Using other types of quantum states to accomplish QSS has already
been well studied in literatures \cite{qi103,qi46}. What we focus in
this paper is the interesting question raised by Refs.
\cite{qi474,qi356} whether QSS can be accomplished using quantum
states having the form of Eq. (\ref{Smolin}), which is a typical
example of bound entangled states. In Ref. \cite{qi483}, the
following QSS protocol was studied.

\bigskip

\textit{The original protocol:}

Alice prepares a 4-qubit Smolin state in the form of Eq. (\ref{Smolin}), and
she keeps qubit $A$ to herself, while sending qubit $B$ to Bob, qubit $C$ to
Charlie and qubit $D$ to Diana respectively. Each party then measures an
arbitrary Pauli matrix $\sigma _{i}$ of his/her respective qubit, and
obtains a result $r_{j}\in \{0,1\}$\ ($j=A,B,C,D$). Then all the parties
announce publicly which observable they measured. If all of them measured
the same observable, from Eq. (\ref{Smolin}) it can be seen that their
results always satisfy $r_{A}\oplus r_{B}\oplus r_{C}\oplus r_{D}=0$ ($%
\oplus $ means addition modulo $2$). Therefore, all the three parties, Bob,
Charlie, and Diana, together can reconstruct Alice's secret bit $r_{A}$.

\bigskip

It was proven in Ref. \cite{qi474} that such a protocol would be secure
against the ``passive'' attacks of external eavesdroppers. However, it was
pointed out in Ref. \cite{qi483} that the protocol would be insecure if the
internal participant Bob cheats with the following intercept-resend strategy.

\bigskip

\textit{The cheating strategy:}

Bob intercepts qubits $C$ and $D$ sent to Charlie and Diana respectively by
Alice, and measures them in the Bell basis. This makes the Smolin state Eq. (%
\ref{Smolin}) collapse into a tensor product of two Bell states with the
same form
\begin{equation}
\left| \psi \right\rangle _{ABCD}=\left| \varphi \right\rangle _{AB}\otimes
\left| \varphi \right\rangle _{CD}.  \label{product}
\end{equation}
Here $\left| \varphi \right\rangle $ is one of the four Bell states
$\left| \Phi ^{\pm }\right\rangle $\ and $\left| \Psi ^{\pm
}\right\rangle $, and from the result of his measurement, Bob knows
which Bell state $\left| \varphi \right\rangle $ is. He then resends
the two qubits of such a Bell state to Charlie and Diana
respectively. Since the Smolin state is merely a
mixture of the product states in the form of Eq. (\ref{product}) where $%
\left| \varphi \right\rangle $ covers all the four possible choices $\left|
\Phi ^{\pm }\right\rangle $\ and $\left| \Psi ^{\pm }\right\rangle $, the
states owned by Alice, Charlie and Diana in this case show no difference
from those in the honest protocol. But since qubit $B$\ owned by Bob is
directly correlated with Alice's qubit $A$ now, Bob alone can know Alice's
secret bit $r_{A}$ when they measure the same observable, without the help
of Charlie and Diana.

\bigskip

This strategy is not only adoptable by Bob. For example, consider that
Charlie intercepts qubits $B$\ and $D$\ and measures them in the Bell basis.
Note that when swapping the position of two of the qubits (e.g., $B$\ and $C$%
), $\left| \Phi ^{\pm }\right\rangle _{AB}\otimes \left| \Phi ^{\pm
}\right\rangle _{CD}$ can be rewritten as
\begin{eqnarray}
\left| \Phi ^{\pm }\right\rangle _{AB}\otimes \left| \Phi ^{\pm
}\right\rangle _{CD} &\longrightarrow &\frac{1}{2}(\left| \Phi
^{+}\right\rangle _{AC}\otimes \left| \Phi ^{+}\right\rangle _{BD}  \nonumber
\\
&&+\left| \Phi ^{-}\right\rangle _{AC}\otimes \left| \Phi ^{-}\right\rangle
_{BD}  \nonumber \\
&&\pm \left| \Psi ^{+}\right\rangle _{AC}\otimes \left| \Psi
^{+}\right\rangle _{BD}  \nonumber \\
&&\pm \left| \Psi ^{-}\right\rangle _{AC}\otimes \left| \Psi
^{-}\right\rangle _{BD}).
\end{eqnarray}
Similar expression can also be found for $\left| \Psi ^{\pm }\right\rangle
_{AB}\otimes \left| \Psi ^{\pm }\right\rangle _{CD}$. Therefore, after
Charlie's measurement, the Smolin state Eq. (\ref{Smolin}) will collapse
into
\begin{equation}
\left| \psi \right\rangle _{ACBD}=\left| \varphi \right\rangle _{AC}\otimes
\left| \varphi \right\rangle _{BD},
\end{equation}
where $\left| \varphi \right\rangle $ is one of the four Bell states $\left|
\Phi ^{\pm }\right\rangle $\ and $\left| \Psi ^{\pm }\right\rangle $.
Comparing with Eq. (\ref{product}), we can see that Charlie can cheat with
the same strategy. So does Diana.

\section{Simplified cheating strategy}

Defeating this cheating alone is easy. Since it requires the cheater to
perform joint measurement on the qubits of the other two parties, we can
restrict Alice to send the qubits one at a time. That is, she does not send
the qubit to the next party until the receipt of the qubit sent to the last
party was confirmed. With this method, the cheater can never have the qubits
of the other two parties simultaneously. Thus he cannot perform joint
measurement on them and the strategy is defeated.

Nevertheless, we would like to pinpoint out that there is an even more
simple cheating strategy which does not require any joint measurement. The
cheater can simply intercept every qubit and measure the same observable of
them (including his own one). Then he resends the measured qubits to the
corresponding parties. As a result, if Alice also measured the same
observable of her qubit, the cheater can infer her result since he has
measured all the other three qubits. Else if Alice measured a different
observable, the result of the four qubits will not have any correlation so
that the cheating will not be detected. Since this strategy involves
individual measurement only, it could be successful even if Alice sends the
qubits one at a time.

\section{Our protocol}

If our purpose is merely to achieve secure QSS, it is not difficult to
defeat all the above cheating strategies. For example, Alice can also
prepare some qubits in pure states. She mixes some of these qubits with
qubit $B$ ($C$ or $D$) and sends them to Bob (Charlie or Diana). By
requiring the other parties to announce their measurement result on some of
these pure states, she can easily check whether there is intercept-resend
attacks on the quantum communication channel between her and each of the
other three parties. After all three quantum channels are verified secure,
she tells the other three parties which qubits are $B$, $C$ and $D$, then
they can accomplish the task of secret sharing with these qubits as
described in the original protocol. Alternatively, Alice can prepare many
copies of Smolin states. She keeps qubits $A$, $B$ and $C$ of each copy to
herself, and sends qubit $D$ to one of the other parties. By measuring $A$
and $B$ in the Bell basis, she can collapse $C$ and $D$\ into a Bell state.
With the Bell state, she can set up a secret key with each of the other
parties with the well-known quantum key distribution protocol \cite{qi10}.
Then the sharing of her secret data can easily be achieved with these secret
keys.

However, these methods cannot help to answer the question whether Smolin
states and bound entanglement can lead to secure QSS. This is because when
pure states are involved, or one party owns more than one qubit of a Smolin
state, the correlation shared between the parties is no longer pure bound
entanglement. Therefore, it is important to study whether a secure QSS
protocol can be built in the framework where only Smolin states are used,
and each party can have one qubit of each copy of Smolin state only, i.e.,
the honest operation on Smolin states must be local operations on single
qubit rather than joint ones on many qubits. Here we purpose such a exotic
protocol.

\bigskip

\textit{Our secure protocol:}

(1) Alice prepares $n$ copies of the 4-qubit Smolin state in the form of Eq.
(\ref{Smolin}). She keeps qubit $A_{j}$ of the $j$th copy ($j=\{1,...,n\}$)
to herself, while sending qubits $B_{j}$ to Bob, qubits $C_{j}$ to Charlie
and qubits $D_{j}$ to Diana ($j=\{1,...,n\}$) respectively. But different
from the original protocol, the order of the qubits sent to each party
should be random. That is, the qubit sequence received by Bob, for example,\
can be $B_{3}B_{6}B_{5}B_{11}B_{4}...$, while that of Charlie and Diana can
be $C_{4}C_{2}C_{9}C_{7}C_{5}...$\ and $D_{4}D_{20}D_{7}D_{3}D_{1}...$\
respectively. The order should be kept secret by Alice herself. Also, each
qubit should be sent only after the receipt of the previous one is confirmed
by the corresponding party.

(2) Alice tells the other three parties which observable to measure for each
of their qubit. She should guarantee that the same observable are measured
for the four qubits of the same copy. But which qubits belong to the same
copy should still be kept secret.

(3) Alice randomly chooses some qubits for the security check. For these
qubits, she asks the other three parties to announce the result of their
measurement, and checks whether $r_{A_{j}}\oplus r_{B_{j}}\oplus
r_{C_{j}}\oplus r_{D_{j}}=0$ is satisfied whenever $A_{j}$, $B_{j}$, $C_{j}$
and $D_{j}$ belonging to the same copy are chosen for the check.

(4) If no disagreed result is found, Alice randomly picks one of the
remaining unchecked copy (suppose that it is the $k$th copy) for secret
sharing. She tells the other three parties the position of qubits $B_{k}$, $%
C_{k}$ and $D_{k}$, so that all the other three parties together can
reconstruct Alice's secret bit $r_{A_{k}}$ for this copy from the equation $%
r_{A_{k}}\oplus r_{B_{k}}\oplus r_{C_{k}}\oplus r_{D_{k}}=0$.

\bigskip

Now we show that the following three important features together
make our protocol secure against known cheating strategies: (i) the
randomness in the secret order of the qubits being sent; (ii) each
qubit is sent only after the receipt of the previous one is
confirmed; (iii) it is decided by Alice which observable the other
parties should measure, and it is not announced until the receipt of
all qubits is confirmed.

Let us consider the most severe case where the number of cheaters is
as large as possible. As stated above, Alice is always assumed to be
honest in QSS. Now if all the other three parties are cheaters, then
they can surely obtain the secret data because any secret sharing
protocol allows the secret to be retrievable when all the three
parties collaborate, even without cheating. Therefore it is natural
to assume that there are two cheaters at the most. For concreteness
and without loss of generality, here we study the case where Diana
is honest while both Bob and Charlie are cheaters and they can
perform any kind of communication (either classical or quantum) with
each other. In fact, due to the symmetric form of Smolin states, the
same security analysis on this case can also apply to the cases
where the two cheaters are Bob and Diana, or Charlie and Diana.
Also, note that external eavesdroppers have less advantages than
internal cheaters since they can attack the quantum communication
channel only, while cannot alter the announcement sent via the
classical channel to cover their attacks. Thus if a QSS protocol is
proven secure against internal cheaters, it is also secure against
external eavesdroppers. Therefore, the case studied here is
sufficient for the security proof.

Let us formulate the model of the cheating strategy of the cheaters
Bob and Charlie. Suppose that they intercepted a qubit being sent to
Diana. Due to the features (ii) of our protocol, they must decide
immediately what kind of qubit should be resent to Diana. There can
be four choices: (a) resend the intercepted qubit intact to Diana;
(b) perform an operation (including projecting the state into a
certain basis, performing an unitary transformation, or making it
entangled with other systems, etc.) on the intercepted qubit, and
then send it to Diana; (c) Prepare another qubit, which may even
entangled with other systems kept by Bob and Charlie, and send it to
Diana; and (d) send Diana another qubit which was sent to Bob or
Charlie, or is previously sent to Diana but intercepted by Bob and
Charlie. Choice (a) is obviously not a cheating anymore. Meanwhile,
all the other choices can be summarized as: Bob and Charlie prepare
the following system
\begin{equation}
\left| bc\otimes d\right\rangle =\sum_{i}\left| \beta _{i}\right\rangle
_{bc}\otimes \left| \gamma _{i}\right\rangle _{d}.  \label{system}
\end{equation}
Here system $d$ is the qubit they will resend to Diana, while system $bc$
can be the system kept at their side and the environment, and may even
include the systems of Alice's and Diana's in choices (b) and (d), and $i$
covers all possible states of these systems. Note that if they measure the
original qubit and then send Diana the resultant state in choice (b), then
system $d$ is in a pure state that does not entangled with system $bc$,
which is simply a special case of Eq. (\ref{system}).

After Diana receives the qubit, due to feature (iii) of our
protocol, the cheaters cannot control the result of Diana's
measurement. Since the qubit is not the original one, Diana's result
does not always show correlation of Smolin states. To avoid the
uncorrelated result from being detected by Alice, the only method
left for the cheaters is to adjust their own announcement in step
(3) of the protocol so that their result looks to be correlated with
that of Diana's. Indeed, after step (2) they can know what result
should have been found by Diana by measuring the original qubit they
intercepted, and it is also possible for them to know the actual
result of Diana's measurement by properly measuring the system $bc$
(if it is completely kept at their side) after they monitor Alice's
announcement to Diana in step (2). However, when they need to
determine their announcement in step (3), Alice has not announced
the ordering of the qubits (i.e., which qubits belong to the same
copy of Smolin state) yet. Note that it is insufficient for Bob and
Charlie to obtain information on this ordering by comparing the
measurement directions Alice announced in step (2) either. This is
because totally there are only three measurement direction
(corresponding to the three Pauli matrices), while the number of
copies of Smolin state is large. Consequently, there will be a large
number of qubits which do not belong to the same copy of Smolin
state, while the measurement directions listed by Alice are the
same. As the number of copies of Smolin state used in the protocol
increases, the amount of mutual information on the ordering Bob and
Charlie gain by comparing the measurement directions will drop
exponentially to zero. Therefore, for the qubits chosen for the
security check, feature (i) ensures that Bob and Charlie do not know
which announcement of their own should be adjusted. Then for any
single copy of Smolin state chosen for the security check in step
(3), the cheaters stand a non-trivial probability (denoted as
$\varepsilon $) to make an inconsistent announcement. The total
probability for the cheaters to escape the detection will be at the
order of $(1-\varepsilon )^{m}$, where $m$ is the number of copies
of Smolin state to which Bob and Charlie apply the attack. This
probability drops exponentially to zero as $m$ increases, so the
cheating will inevitably be detected if $m$ is large. On the other
hand, if $m$ is small, the probability for these $m$ copies of
Smolin state to be chosen as the $k$th copy for the final secret
sharing in step (4) will drop to zero as the total number of copies
of Smolin state used in the protocol increases, so that the cheating
is fruitless. Thus it is proven that our protocol is secure against
the cheating strategy above.

It is important to note that in our protocol, after Alice performs
permutation on all Smolin states, the resultant states are still
bound-entangled. The reason is that the permutation operation can in
fact be viewed as a local operation, because in step (1) of the
protocol, the qubit sequence received by Bob is merely the
permutation of all $B_{j}$'s (e.g.,
$B_{3}B_{6}B_{5}B_{11}B_{4}...$), while
that of Charlie and Diana are all $%
C_{j}$'s and $D_{j}$'s respectively. There is no joint operation
between the qubits $A$, $B$, $C$ and $D$. That is, suppose that the
four participants share many copies of Smolin state, each
participant has one qubit from each copy. Then the above permutation
can be accomplished by each participants locally. It is a known fact
that Smolin states cannot be distilled to pure entangled form with
LOCC. Therefore the resultant states still cannot be distilled with
LOCC either, thus it still satisfy the definition of bound entangled
states. For this reason, what we achieved here is not merely another
QSS protocol. The significance of our result is that the QSS
protocol proposed here is based on bound entangled state alone.

We have to point out that we currently cannot prove the generality
of the model of the cheating strategy we studied above, because
there may potentially exist strategies which are beyond our current
imagination. Therefore, whether our specific protocol is
unconditionally secure against any cheating strategy or not is still
an open question. Nevertheless, the above model seems to cover all
attacks currently known. Therefore, before a different cheating
strategy fell outside the above model could be found in the future,
our result seems to give a positive answer to the question whether
Smolin states alone can lead to secure QSS. This is in contrast to
the conclusion of Ref. \cite{qi483}.

It also seems that generalized Smolin states \cite{qi474} can lead to secure
quantum secret sharing between more participants too. Here the generalized
Smolin states mean the $2n$-qubit ($n>2$) bound entangled states defined as
follows. Let $U_{n}^{(m)}=I^{\otimes n-1}\otimes \sigma _{m}$ ($m=0,1,2,3,$ $%
n=1,2,3,...$) be a class of unitary operations, where $\sigma _{0}=I$\ is
the identity acting on the $2$-dimensional Hilbert space $C^{2}$ and $\sigma
_{i}$ ($i=0,1,2,3$) are the standard Pauli matrices. Let $\rho _{2}=\left|
\Psi ^{-}\right\rangle \left\langle \Psi ^{-}\right| $, and denote the
density matrix of the original 4-qubit Smolin state (Eq. (\ref{Smolin})) as $%
\rho _{4}$. Then the density matrix of $2n$-qubit ($n>2$) generalized Smolin
state is
\begin{equation}
\rho _{2n}=\frac{1}{4}\sum\limits_{m=0}^{3}U_{2(n-1)}^{(m)}\rho
_{2(n-1)}U_{2(n-1)}^{(m)}\otimes U_{2}^{(m)}\rho _{2}U_{2}^{(m)}.
\end{equation}
(Please see Ref. \cite{qi474} for details.) When the state is shared by $2n$
parties (each party has one qubit) and they measure the same observable,
their results will always satisfy $\sum\nolimits_{j=1}^{2n}\oplus r_{j}=0$.
Therefore we can see that a secure quantum secret sharing between Alice and
other $(2n-1)$ parties can be accomplished with a protocol similar to our
above secure protocol with original Smolin states, by including the
following main features. (i) Alice prepares many copies of the $2n$-qubit
generalized Smolin state, and sends them to the other parties in random
order. It is not announced which qubits belong to the same copy, until all
secure checks are successfully finished. (ii) Each qubit is sent only after
the receipt of the previous one is confirmed. (iii) It is decided by Alice
which observable the other parties should measure, and it is not announced
until the receipt of all qubits is confirmed.

\section{Summary and discussions}

Thus we proposed a QSS protocol secure against known cheating
strategies. We would like to emphasize that the present protocol is,
to our best knowledge, the first example of secure QSS in terms of
bound entangled states alone. This result suggests an positive
answer to the question in Refs. \cite {qi474,qi356} whether Smolin
states can lead to secure QSS. As to the more general question in
Refs. \cite{qi356} whether there are cases when violation of local
realism is necessary but not sufficient condition for QSS, our
result seems to suggest that we need not to search for such cases in
the framework of the original and generalized Smolin states.

Our protocol is also feasible for practical implementation. At the first
glance, there seems to have a difficulty since the qubits received by Bob,
Charlie and Diana in step (1) need to be kept unmeasured until Alice
announces which observable to measure in step (2). To this date, keeping a
quantum state for a long period of time is still a technical challenge.
Nevertheless, in practice Alice can uses the well-known quantum key
distribution protocol (e.g., Ref. \cite{qi10}) to setup a secret string with
each of the other parties beforehand, so that she can tell him secretly
which observable he is to measure. Then delaying the measurement is no
longer necessary. Therefore our protocol can be implemented as long as a
source of Smolin states is available. Though in this case, not merely bound
entangled states are used in the protocol, it is made simple to realize
secure QSS with state-of-the-art technology.

Finally, we would like to propose a feasible scheme to prepare Smolin states
experimentally \cite{qi474}. The quantum circuit for this scheme is shown in
Fig.1. The input part contains six qubits, in which the ancillary qubits $%
\alpha$ and $\beta$ are initialized in state $\left| ++\right\rangle
_{\alpha\beta}$ (here, $\left| +\right\rangle =(\left| 0\right\rangle
+\left| 1\right\rangle )/\sqrt{2}$) and the target qubits in the tensor
product of two Bell states $\left| \Phi ^{+}\right\rangle _{AB}\otimes
\left| \Phi ^{+}\right\rangle _{CD}$. First, let $\beta$ be the control
qubit and perform the controlled-$\sigma _{z}$ operations on qubits $\beta A$
and $\beta C$, respectively. Then, let $\alpha$ be the control qubit and
perform the controlled-$\sigma _{x}$ operations on qubits $\alpha B$ and $%
\alpha D$, respectively. This procedure of the target qubits can be
formulated by
\begin{equation}
\rho _{ABCD}^{S}=\mathtt{Tr}_{\alpha\beta}[U\rho _{in}U^{\dagger }],
\label{1}
\end{equation}
where the input state is $\rho _{in}=\left| \psi \right\rangle
_{in}\left\langle \psi \right| $ with $\left| \psi \right\rangle
_{in}=\left| ++\right\rangle _{\alpha\beta}\otimes \left| \Phi
^{+}\right\rangle _{AB}\otimes \left| \Phi ^{+}\right\rangle _{CD}$, and the
unitary transformation takes the form
\begin{eqnarray}
U &=&\left| 00\right\rangle _{\alpha\beta}\left\langle 00\right| \otimes
I_{ABCD}  \nonumber \\
&&+\left| 01\right\rangle _{\alpha\beta}\left\langle 01\right| \otimes
\sigma _{z}^{A}\sigma _{z}^{C}I_{BD}  \nonumber \\
&&+\left| 10\right\rangle _{\alpha\beta}\left\langle 10\right| \otimes
I_{AC}\sigma _{x}^{B}\sigma _{x}^{D}  \nonumber \\
&&+\left| 11\right\rangle _{\alpha\beta}\left\langle 11\right| \otimes
\sigma _{z}^{A}\sigma _{x}^{B}\sigma _{z}^{C}\sigma _{x}^{D}.
\end{eqnarray}
After this procedure, the output state will be the desired quantum state,
\emph{i.e.}, the Smolin state $\rho _{ABCD}^{S}$. Therefore with any source
of Bell states currently available, Smolin states may be generated with this
scenario, and thus our protocol is expected to be implemented in principle
in near future.

\begin{figure}[h]
\includegraphics{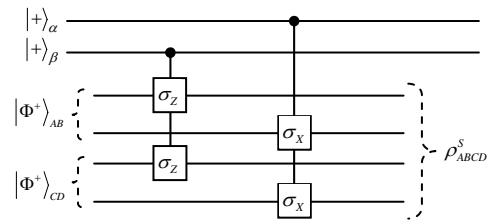}
\caption{\label{fig:epsart}The quantum circuit for generating the
Smolin state.}
\end{figure}

The work was supported by the NSFC under grant Nos.10605041 and
10429401, the RGC grant of Hong Kong, the NSF of Guangdong province
under Grant No.06023145, and the Foundation of Zhongshan University
Advanced Research Center.



\end{document}